# Crystalline 'Genes' in Metallic Liquids

Yang Sun[1,2 †], Feng Zhang[2 †], Zhuo Ye[2], Xiaowei Fang[1,2], Zejun Ding[1], Cai-Zhuang Wang[2,3], Mikhail I. Mendelev[2], Ryan T. Ott[2], M. J. Kramer[2], Kai-Ming Ho[1,2,3 *]

[1]Hefei National Laboratory for Physical Sciences at the Microscale and Department of Physics, University of Science and Technology of China, Hefei, Anhui 230026, China

[2]Ames Laboratory, US Department of Energy, Ames, Iowa 50011, USA

[3]Department of Physics, Iowa State University, Ames, Iowa 50011, USA

[†] These authors contributed equally to this work.
[*] Email: kmh@ameslab.gov

**Abstract**

**The underlying structural order that transcends the liquid, glass and crystalline states is identified using an efficient genetic algorithm (GA). GA identifies the most common energetically favorable packing motif in crystalline structures close to the alloy's Al-10 at.% Sm composition. These motifs are in turn compared to the observed packing motifs in the actual liquid structures using a cluster-alignment method which reveals the average topology. Conventional descriptions of the short-range order, such as Voronoi tessellation, are too rigid in their analysis of the configurational poly-types when describing the chemical and topological ordering during transition from undercooled metallic liquids to crystalline phases or glass. Our approach here brings new insight into describing mesoscopic order-disorder transitions in condensed matter physics.**

1. INTRODUCTION

While the general consensus is that noncrystalline metallic phases (i.e. liquids and glasses) have clear elements of short- and medium-range order[1, 2, 3, 4, 5, 6, 7, 8], they are not easily identified due to the amorphicity in the structures. From an energetic standpoint, this difficulty severely limits our ability to acquire an accurate description of order parameters required for extending the traditional models of Gibbs free energy and associated analyses to understand and predict phase transformations. In this Letter, we establish a new order-mining approach in amorphous systems which examines the underlying short-range topological similarities between amorphous and crystalline structures. Using the Al-Sm system as an example, we demonstrate that composition-specific short-range order (SRO) dominates both amorphous and crystalline phases during amorphization and devitrification; and thus plays a key role in the phase selection during these processes.

Various methods have been proposed to identify the system-specific SRO. Models based on random dense packing of hard spheres use differential component sizes as the deterministic factor for local atomic packing[9], which has proven to be insufficient in various cases[10]. Methods for directly extracting local orders out of atomistic simulations are also available, such as Voronoi tessellation[11] and Honeycutt-Anderson (HA) common neighbor analysis[12]. The Voronoi tessellation method is too sensitive to geometric distortions of the clusters, especially when the coordination number of the cluster is high[3, 13]. The HA indexing method, while generally robust against geometric distortions, only captures a fragment of a complete cluster; and thus sometimes fails to distinguish different clusters sharing common partial motifs, such as the icosahedron cluster and the C16 prototype for $AB_2$ compounds[14].

Alternative methods use pre-selected prototype templates to calibrate the local structural order, based on certain parameters that can distinguish different templates. Examples are the bond-orientational order analysis[15] and the cluster alignment method[14]. In the bond-orientational order analysis, a set of rotation-invariant parameters are constructed to describe the collective orientational order of the atomic bonds surrounding a central atom[15]. These parameters are generally sensitive to different symmetries around the central atom, and thus can effectively distinguish commonly seen packing motifs in metallic systems, including icosahedron, fcc, hcp, etc. The cluster alignment method is analogous to the structural alignment used to identify regions of similarity in biomolecules[16], and is designed to quantify the similarity between a cluster extracted from amorphous samples and an ideal template[14]. To apply such template-assisted methods, it is important to choose a complete set of templates. For metallic systems, the local packing geometry usually shows a strong dependence on the overall composition. In the following, we demonstrate an approach for establishing the composition-specific templates.

No structural motifs are exclusive to amorphous structures; that is, they *can* pack into crystals. This is true even for the icosahedral cluster commonly used to model the SRO in undercooled metallic liquids or glasses[17, 1817, 1817, 18]: Although the icosahedral cluster contains non-crystallographic five-fold symmetry, many crystals contain local icosahedral ordering[19]. Amorphicity appears when nucleation and growth of such clusters is inhibited under certain processing conditions. Therefore, a global search of crystalline structures could yield potential packing templates that can characterize amorphous structures of similar compositions. Given the preference of such templates, the crystalline structures containing them should have relatively low energies. We use a genetic algorithm (GA)[20], which is a robust means to locate low-energy configurations in crystals[21], to perform the global search in order to identify novel template motifs specific to a certain composition range. Then we use the cluster alignment method to check the preference of the GA-identified motifs in real undercooled liquids.

The Al-Sm example is used for this study. With ~ 10 at.% Sm, the Al-Sm system can be rapidly quenched to an amorphous state which, in turn, shows a rich sequence of meta-stable states upon heating[22]. The larger Sm atom cannot fit into the small fcc Al lattice thus inhibiting crystallization of the dominant thermodynamically stable phase. We postulate the key to controlling the transition from liquid to glass lies in understanding the chemical and topological SRO around the solute Sm atoms. In turn, the phase evolution during the devitrification process is mediated by the local rearrangements of the short to medium-range order (MRO) upon heating.

## 2. METHODS

**Adaptive Genetic algorithm.** A classical potential in the Finnis-Sinclair[23] form is employed to quickly calculate energy during the GA search. A portion of low-energy structures obtained in the GA search are then collected for accurate calculations using the density functional theory (DFT). The DFT results are used to adapt the parameters of the auxiliary classical potential. The above process is repeated until the structures collected in the DFT calculation pool are converged.

*Ab initio* **molecular dynamics.** The constant number of atoms, volume and temperature (*NVT*) ensemble is applied with Nose-Hoover thermostats. The Verlet algorithm is used to integrate Newton's equation of motion, using a time step of 3 fs. Three different samples, all with 450 Al atoms and 50 Sm atoms, are created independently for better statistical analysis. To create these samples, randomly generated configurations with cubic supercell are equilibrated at 2100 K over 2000 time steps. Then each sample is cooled down to 800 K, well-below the melting temperature 1200K[24] with a cooling rate of 2.2×10$^{13}$ K/s. After that, the structures at 1300 K, 1000 K and 800 K are collected separately for further isothermal annealing for about 6,000 time steps. The first 3,000 time steps are not used in the analysis to ensure equilibrium has been reached.

**Density functional theory (DFT).** All DFT calculations are performed using the Vienna *ab initio* simulation package (VASP)[25]. The projected augmented-wave (PAW) method is used to describe the electron-ion interaction, and the generalized gradient approximation (GGA) in the Predew-Burke-Ernzerhof (PBE) form is employed for the exchange-correlation energy functional.

**X-ray diffraction measurements.** The Al-10at.% Sm alloy was synthesized by arc melting high-purity elements (>99.9%) in a Ti-gettered Ar environment. After melting and flipping several time to assure homogeneity, the alloy was cast into a 6 mm mold. The cast ingot was then melted in an induction furnace and injection cast into a 1.6 mm mold to form rods that were placed in silica capillary tubes. The X-ray experiments were performed at Sector 6-ID-D of the Advanced Photon Source at Argonne National Laboratory. The samples were heated in furnace that was placed 347 mm upstream from a MAR CCD detector that was positioned off-axis to collect a higher $Q$-range. X-rays with an energy of 100 keV ($\lambda$ = 0.1245 A) were utilized in the experiments, which were performed in transmission mode. The collected diffraction patterns were integrated using Fit2D software[26], and corrected for absorption, polarization, multiple scattering, and Compton scattering[27]. The total scattering functions, *S(Q)*, at the different temperatures were calculated according to

$$S(Q) = 1 + \frac{\left[ I(Q) - \sum_{i=1}^{n} a_i |f_i(Q)|^2 \right]}{\left| \sum_{i=1}^{n} a_i f_i(Q) \right|^2}$$

Where $Q = 4\pi \sin\theta/\lambda$, $I(Q)$ is the coherently scattered portion of the total intensity, $a_i$ is the atomic proportion of each element, and $f_i(Q)$ is the $Q$-dependent scattering factor for each element.

## 3. RESULTS AND DISCUSSION

As shown in Fig. 1a, the Al-Sm system has already shown a rich collection of Sm-centered ordering in known crystalline compounds. Since the composition of the target system (~10 at.% Sm) is different from any of the compounds shown in Fig. 1a, we expect to see new structural motifs characterizing this composition range that are not covered in Fig. 1a.

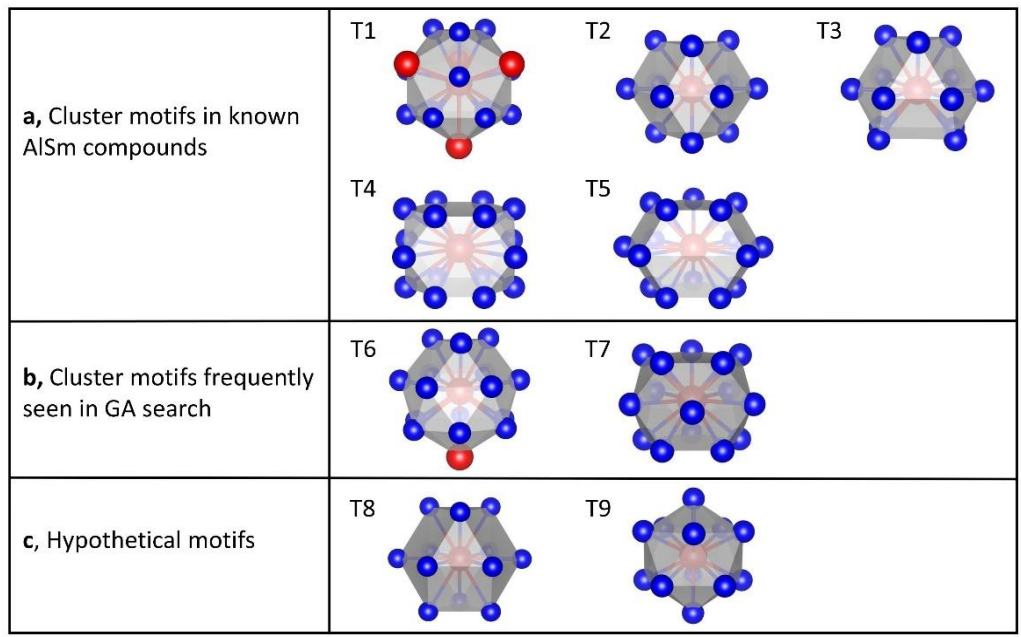

**Figure 1 | Sm-centered template cluster motifs containing the first atomic shell.** Red stands for Sm and blue for Al. **a**, template motifs extracted from known Al-Sm crystalline compounds[28]: T1 is extracted from the Al$_2$Sm phase, a typical C.N. 16

Frank-Kasper polyhedron; T2 from $Al_3Sm$; T3 from $γ$-$Al_4Sm$; T4 from $β$-$Al_4Sm$ and $α$-$Al_{11}Sm_3$; T5 is another Sm-centered motif in $α$-$Al_{11}Sm_3$. **b**, two motifs frequently appearing in new crystalline structures found by GA search. **c**, two additional hypothetical motifs: T8 is the building block for fcc structures; T9 is an icosahedron.

To identify the missing motifs, we have performed GA searches for low-energy crystal structures with unit cells containing up to 50 atoms and a narrow Sm composition range between 0.1 and 0.125. Indeed, two new Sm-centered motifs, T6 and T7, as shown in Fig. 1b, are frequently seen in relatively low-energy structures found in the GA search. The first shell of the T6 motif consists of a top triangular Al layer followed by two hexagonal Al layers and a bottom Sm atom, whereas the first shell of T7 contains three successive pentagonal Al layers. The formation energy ($E_{form}$) of the structures containing T6 or T7 motifs is plotted in Fig. 2 as a function of $x_{Sm}$. $E_{form}$ is referenced to the stable $Al_3Sm$ phase and fcc Al. The positive values of $E_{form}$ show that these structures are unstable with respect to separation into Al and $Al_3Sm$ ground-state structures, consistent with the fact that the Al-Sm phase diagram shows no stable Al-richer compounds than $Al_3Sm$[24]. However, under fast quenching conditions, the pathway to phase separation into the equilibrium mixture of Al and $Al_3Sm$ can be kinetically by-passed, and local clusters T6 or T7 can still be formed.

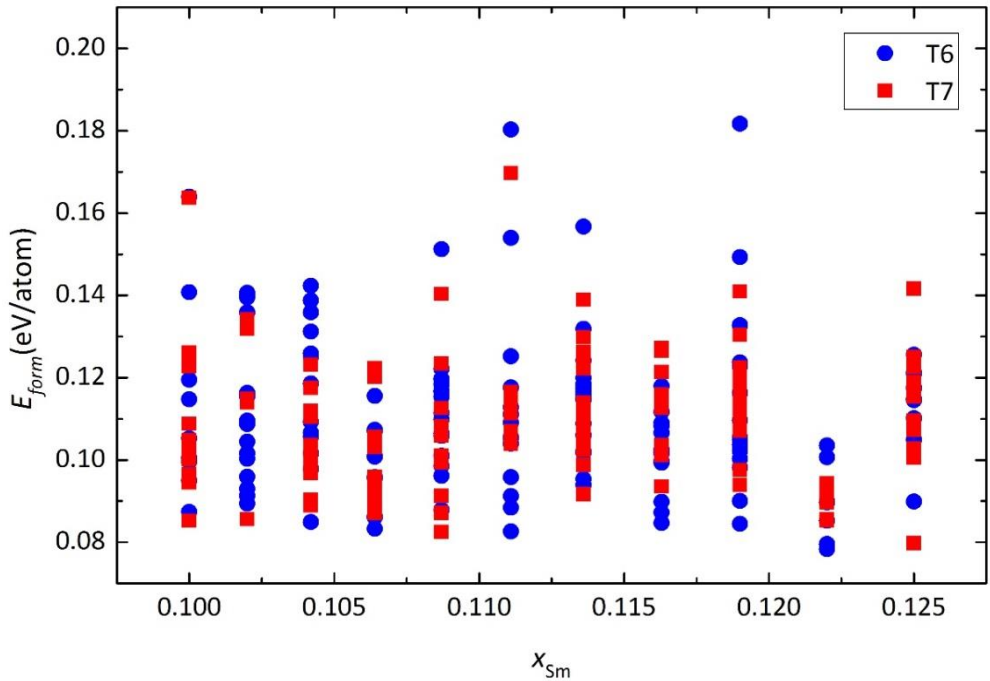

**Figure 2 | Formation energy as a function of the Sm composition, for a series of new phases found in the GA search that contain T6 or T7 motifs.**

The preference of T6 and T7 will be checked against various other clusters in undercooled $Al_{90}Sm_{10}$ liquids. In addition to those existing in known crystalline compounds [T1-T5 in Fig. 1a], we introduce two more hypothetical competitors: T8 and T9 as shown in Fig. 1c. T8 is the building block of the fcc structure for pure Al. T9 represents the icosahedral SRO commonly seen in amorphous structures.

To generate undercooled $Al_{90}Sm_{10}$ samples, *ab initio* molecular dynamics (AIMD) simulations are performed on the $Al_{90}Sm_{10}$ system at various temperatures. The unit cell contains 500 atoms, large enough to fully accommodate the SRO. In Fig. 3a-c, the calculated partial pair correlation functions (PPCFs) of the samples are presented at various temperatures. The PPCF is averaged over the three independently prepared samples. An "error band" is included by sweeping the error bar across all positions.

The error band for the Al-Al and Al-Sm PPCFs is vanishingly narrow. For the Sm-Sm PPCF, the error band is slightly broadened, since Sm is the sparse species in the system. Overall, Fig. 3a-c shows that the PPCFs for the three different samples are reasonably well converged, and thus the structural features extracted from these samples are statistically valid. Furthermore, we calculated the structure factor $S(q)$ of the simulated samples at 1300K, (details for the calculation is given in Section I of the Supplementary Information). As shown in Fig. 3d, the calculated $S(q)$ compares favorably with that measured in X-ray diffraction experiments, reassuring that our simulations reliably capture the structural properties of the $Al_{90}Sm_{10}$ system.

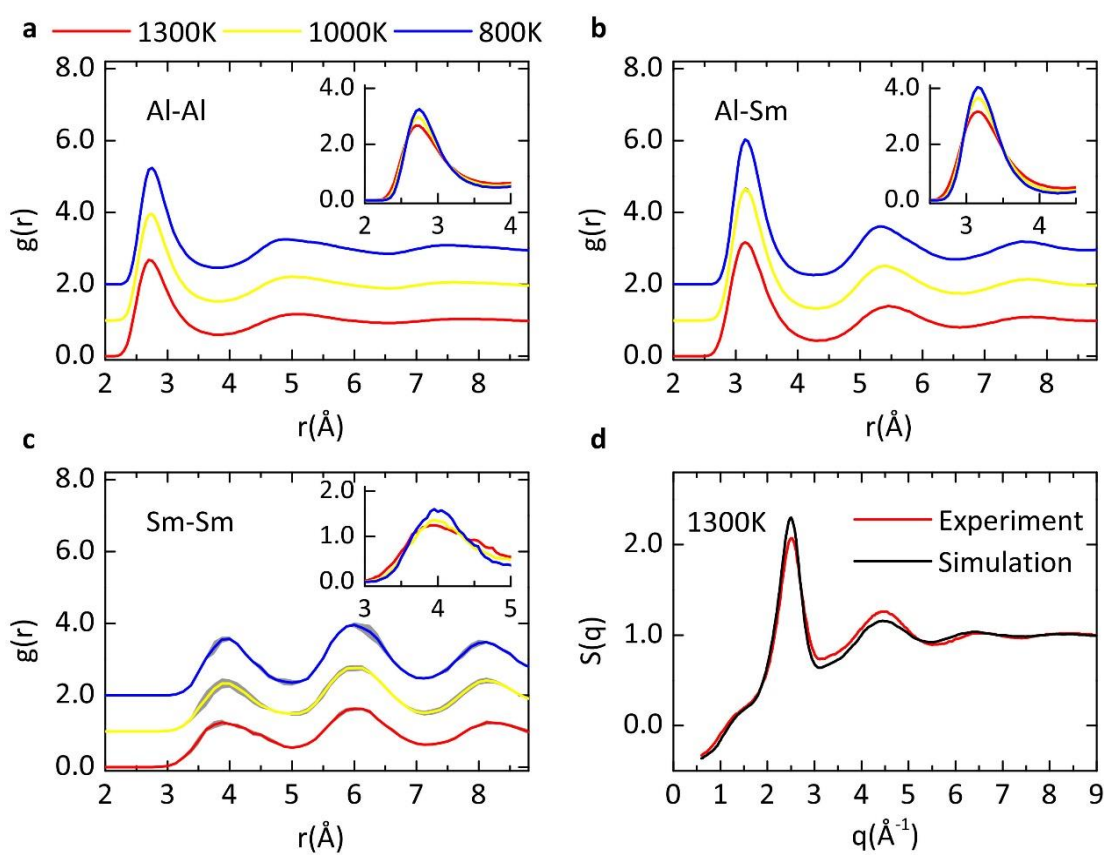

**Figure 3 | The partial pair correlation function (PPCF) and the structure factor for Al90Sm10. a-c**, the PPCF averaged over three AIMD Al450Sm50 samples. The grey error band is generated by sweeping the error bar over all positions. The curves of different temperatures are shifted vertically for clarity. The inserts zoom in the first peak which is enhanced during the cooling process. **d**, the structure factor obtained

both in AIMD simulations and in experiments.

To characterize the SRO surrounding Sm atoms in the amorphous samples created by AIMD, Sm-centered clusters are extracted from the samples, and are aligned against all the cluster templates given in Fig. 1, following the individual template-cluster alignment procedure described in Ref. 14. An alignment score, describing how an as-extracted cluster deviates from a perfect template, is defined as

$$f = \min_{0.80 \leq \alpha \leq 1.2} \left( \frac{1}{N} \sum_{i=1}^{N} \frac{(\mathbf{r}_{ic} - \alpha \mathbf{r}_{it})^2}{(\alpha \mathbf{r}_{it})^2} \right)^{1/2}, \quad (1)$$

where $N$ is the number of the neighbor atoms in the template; $\mathbf{r}_{ic}$ and $\mathbf{r}_{it}$ are the atom positions in the aligned cluster and template, respectively; and $\alpha$ is a coeffient to adapt the template's bond length. The smaller the alignment score is, the more similar the cluster is to the template. If the alignment score is larger than a cut-off value of 0.19, then the template fails to characterize the cluster (for more details about the selection of the cut-off value, see Section II of the Supplementary Information).

We first check the similarity among the cluster templates shown in Fig. 1. It turns out that while T4, T5 and T9 are well separated from others, the remaining motifs show certain similarities among themselves. Among this group of similar motifs, the GA-identifed T6 motif is the most representative one and best describes the clusters in the AIMD samples (please refer to Section III of the Supplemental Information for more details). In Fig. 4a, we show the population of four topologically different motifs T4, T5, T6 and T9 in AIMD samples at several temperatures, using T6 to represent the group with similarities. The temperature varies from above the melting point of 1253 K[24] to deeply undercooled regime. The averaged population over three independent samples is shown along with the error bar. At $T = 1300$ K, less than 10% Sm-centered clusters from the liquid samples can be characterized with the four typical motifs, due to poorly developed local structural order at this temperature. Among them, the T6 motif already has a considerable population of 8%. With the temperature decreasing, the unknown population reduces, indicating the enhanced SRO. The same trend can be seen from the increment of the first peak height in the PPCFs during the cooling process, as shown in the inserts of Fig. 3a-c. At $T = 800$ K, the total population of identified clusters increase to 45%, most of which belongs to the T6 motif. This clearly shows that the T6 motifs, commonly seen in Al-Sm crystalline structures with $x_{Sm}$ close to 0.1, are also characteristic of undercooled amorphous structures with similar compositions, while other three motifs are essentially nonexistent in the samples.

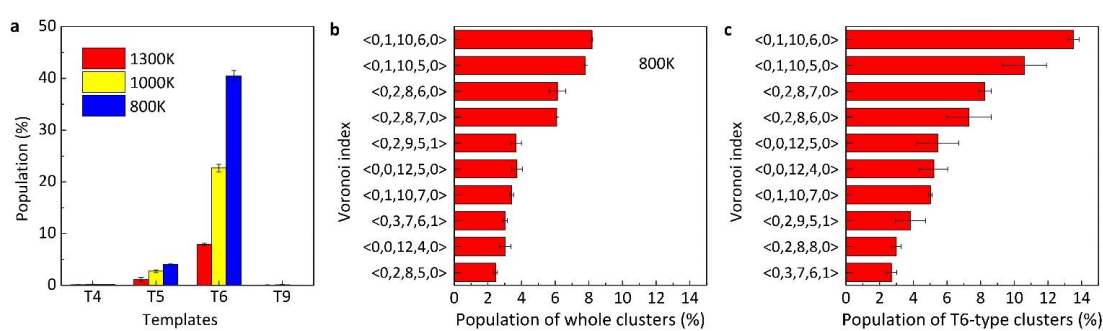

**Figure 4. | The population of SRO identified by template-assisted cluster alignment and by the Voronoi tessellation analysis. a,** populations of four typical cluster motifs in AIMD samples obtained by template-assisted cluster alignment. **b,** Populations of 10 most popular Voronoi polyhedrons surrounding Sm atoms in the AIMD samples at 800K. **c,** populations of 10 most popular Voronoi polyhedrons surrounding the sub-group of Sm atoms characterized as T6-type at 800K shown in **a**. Note that no dominate Voronoi polyhedron can be seen in both **b** and **c**, indicating the Voronoi tessellation is blind with the clusters.

The above characterization of structural ordering in the $Al_{90}Sm_{10}$ system with the help of GA-identified motifs cannot be achieved in the conventional Voronoi tessellation analysis. Fig. 4b gives the 10 most frequent Voronoi polyhedrons surrounding Sm atoms in the amorphous samples at $T = 800$ K. One can see a widespread distribution of these polyhedrons. The most popular polyhedron with an index <0, 1, 10, 6, 0> only has a population of 8%. All these polyhedrons have large coordination numbers of 16 or more. Consequently, small distortions can result in dramatic change of the Voronoi indices. Moreover, it has been demonstrated that Voronoi indexing on clusters that carry close-packing features such as fcc or hcp are particularly vulnerable to distortions[29]. For these reasons, the Voronoi index corresponding to the ideal T6 motif, <0, 3, 12, 1, 0>, only carries a vanishingly small population. In Fig. 4c, we show the distribution of the 10 most frequent Voronoi polyhedrons surrounding a sub-group of Sm atoms that have been characterized to be T6-type. Interestingly, this distribution highly resembles the one for the entire set of Sm atoms shown in Fig. 4b. This demonstrates that in Voronoi tessellation, this sub-set of Sm atoms is treated as if they were randomly selected from the whole sample. In other words, the order characterization achieved by GA and template-assisted cluster alignment is not reflected in the Voronoi tessellation at all.

Experimentally, it has been observed that melt spun amorphous $Al_{90}Sm_{10}$ samples initially transforms into a metastable big cubic phase upon heating to 506K[30]. Recently, we have solved the atomic structure of this big cubic phase. In this structure, approximately 60% of Sm-centered clusters belong to the T6 motif. This similarity of the SRO between the big cubic phase and the amorphous parent structure is probably the reason why the metastable big cubic phase appears first in this complicated multi-step devitrification process.

## 4. Conclusion

In summary, we develop a systematical scheme integrating genetic algorithm, *ab initio* MD simulations and the cluster alignment method to reveal dominant SROs in liquid alloys. We show that the SROs characterizing low-energy crystalline Al-Sm structures with ~10 at.% Sm also have abundant population in deeply undercooled $Al_{90}Sm_{10}$ liquids. Our work provides a systematic approach to address a key question in determining the prevailing forms of non-crystalline order in liquids and glasses. Future work will address the spatial extent, interconnections and stability of regions with strong SRO within the undercooled liquid, and the role these networks play in nucleation and phase selection during phase transformations.

**Acknowledgements**

Work at Ames Laboratory was supported by the US Department of Energy, Basic Energy Sciences, Division of Materials Science and Engineering, under Contract No. DE-AC02-07CH11358, including a grant of computer time at the National Energy Research Supercomputing Center (NERSC) in Berkeley, CA. Y.S acknowledges the support from China Scholarship Council (File No. 201406340015). Z.J.D. acknowledges support from the National Natural Science Foundation of China (No. 11274288) and the National Basic Research Program of China (No. 2011CB932801 and No. 2012CB933702).

# Supplementary Information: Crystalline 'Gene' in Metallic Liquids

Yang Sun[1,2 †], Feng Zhang[2 †], Zhuo Ye[2], Xiaowei Fang[1,2], Zejun Ding[1], Cai-Zhuang Wang[2,3], Mikhail I. Mendelev[2], Ryan T. Ott[2], Matthew J. Kramer[2], Kai-Ming Ho[1,2,3 *]

[1]Hefei National Laboratory for Physical Sciences at the Microscale and Department of Physics, University of Science and Technology of China, Hefei, Anhui 230026, China

[2]Ames Laboratory, US Department of Energy, Ames, Iowa 50011, USA

[3]Department of Physics, Iowa State University, Ames, Iowa 50011, USA

† These authors contributed equally to this work.
* Email: kmh@ameslab.gov

## I. Accurate calculation of structure factor

Traditionally, structure factor (SF), $S(q)$, is calculated from partial pair correlation function (PPCF) $g_{\alpha\beta}(r)$, using the Faber-Ziman formalism [S1] as follows:

$$S_{\alpha\beta}(q) = 1 + 4\pi\rho \int_0^\infty \left[ g_{\alpha\beta}(r) - 1 \right] \frac{\sin qr}{qr} r^2 dr, \qquad (S1)$$

$$S(q) = \sum_{\alpha,\beta} c_\alpha c_\beta f_\alpha f_\beta \left[ S_{\alpha\beta}(q) - 1 \right]. \qquad (S2)$$

Here, $\alpha, \beta$ denotes different atom species, $\rho$ is the atomic density, $c_\alpha$ and $c_\beta$ are the molar fractional compositions of the components, and $f_\alpha$ and $f_\beta$ are the atomic scattering factors. The equations are exact; however, the length of the PPCF from the *ab-initio* molecular dynamics simulations (AIMD) cannot exceed the half of the simulation box size (here, ~ 11 Å). Therefore, the integral in equation (S1) has to be truncated, resulting in transformation ripples at small q value. Thus, the structure factor directly obtained from AIMD will generally feature large inaccuracies at the small q regime. Here, to improve the accuracy of $S(q)$, we applied an elongation algorithm [S2] to extend the PPCFs to larger distances, using large–scale classic molecular dynamics (CMD) simulations with the semi-empirical Finnis-Sinclair (FS) potential [S3].

We first use the AIMD results at 1300K to fit the FS potentials. The fitting process uses AIMD PPCF up to 7.0 Å as the target functions, and includes the lattice parameters and formation energy of AlSm compounds obtained from DFT calculations. The fitting methods are described in detail in Ref. S4,S5,S6. Then the CMD simulations are carried out with 5000 atoms per unit cell, so that the effective length of PPCF from CMD simulations is determined up to ~ 23 Å. Fig. S1 demonstrates that the CMD simulations with FS potential almost exactly reproduce the AIMD PPCF.

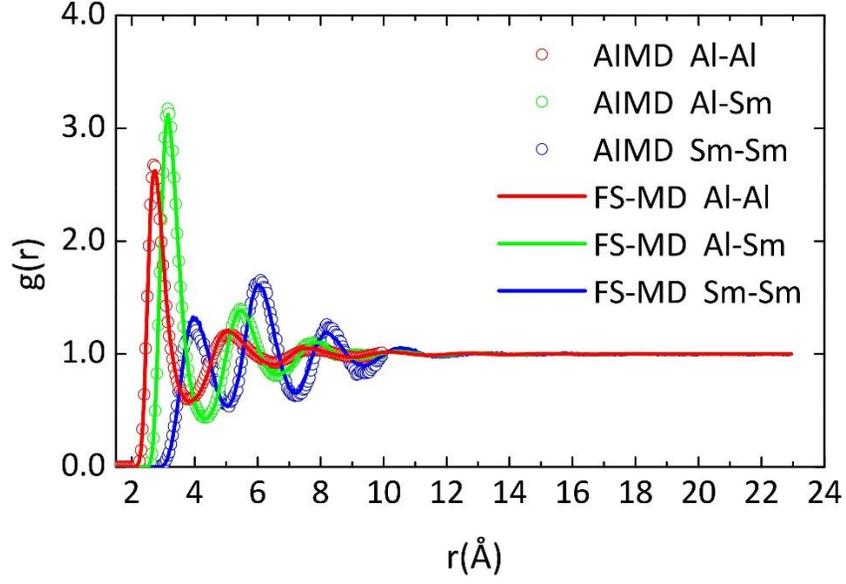

Figure S1 The elongation of partial pair correlation functions from *ab initio* molecular dynamics simulations. The circles shows the PPCFs calculated from AIMD simulations with 500 atoms per unit cell. Solid line shows the PPCFs calculated from CMD simulations using FS potential with 5000 atoms per unit cell.

Finally, to improve the accuracy of the SF, the PPCFs are elongated as:

$$g_{\alpha\beta}^{elong} = \begin{cases} g_{\alpha\beta}^{AIMD}(r), \ r < r_1 \\ g_{\alpha\beta}^{AIMD}(r)\frac{r_2 - r}{r_2 - r_1} + g_{\alpha\beta}^{FS}(r)\frac{r - r_1}{r_2 - r_1}, \ r_1 < r < r_2, \\ g_{\alpha\beta}^{FS}(r), \ r > r_2 \end{cases} \quad (S3)$$

where $r_1$ is 7.0 Å and $r_2$ is 10.0 Å. The final PPCF is set to the AIMD PPCF at $r < r_1$ and the FS PPCF at $r > r_2$. In the interval from $r_1$ to $r_2$, a linear interpolation between the AIMD PPCF and the FS PPCF is used. As shown in the Fig. 3d in the main text, the structure factor calculated using this elongation method compares favorably with that obtained in X-ray diffraction measurements.

## II. Determination of the cut-off value for the alignment score

In order to select an appropriate cut-off alignment score, we first independently aligned the Sm-centered clusters extracted from the AIMD samples against each cluster motif shown in Fig. 1 of the main text. Fig. S2 shows the distribution of the alignment score as defined in equation (1) of the main text for all cluster motifs at three different temperatures, where one can see vastly different peak positions for different cluster motifs. Since the alignment score reflects the similarity of the as-extracted clusters to the template motif, in general, the lower the peak position is, the higher popularity the corresponding cluster motif has in the AIMD samples. Meanwhile, the peak position for each template left-shifts as the temperature decreases, showing enhanced local ordering when the liquid gets more deeply undercooled. Here, in order to determine the population of each temperature in the AIMD samples, we choose a cut-off alignment score of 0.19, which is close to the peak position for the most dominant T6 motif at $T = 800$ K.

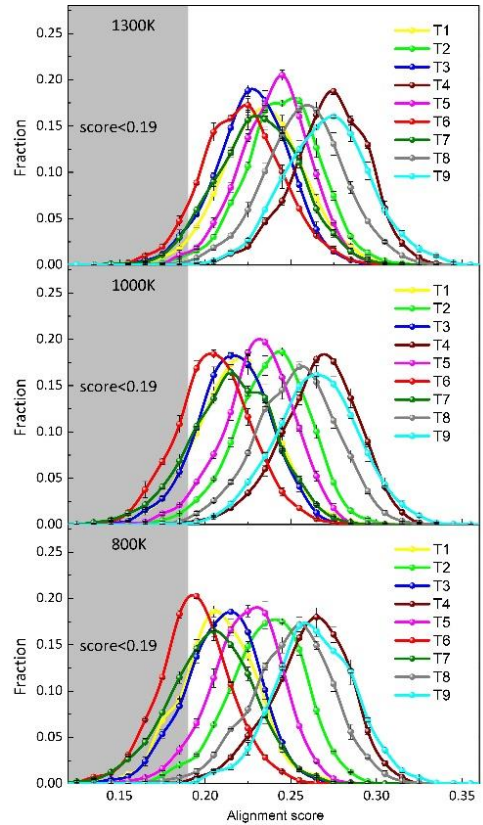

**Figure S2** The alignment score distribution for the template clusters at T = 1300K, 1000K, and 800 K. T1-T9 indexes correspond to the template clusters in Fig. 1 of the main text.

### III. Classification of the template cluster motifs

Cluster alignment algorithm can also be used to check the similarity among the template motifs given in Fig. 1 of the main text. In Fig. S3a, we connect any pair of motifs if the score of their mutual alignment is smaller than the cut-off value of 0.19. While the motifs T4, T5, and T9 are isolated, the remaining cluster motifs show complex correlations among themselves.

After a closer inspection, one can identify some common features of this group: 1) T1 and T6 share a hexagonal pyramid at the bottom as shown in Fig. S3b; 2) T2, T3, T6, T7 and T8 share a "triangle and hexagon" packing of Al atoms, which is characteristic of the A-B stacking in close packing structures as shown in Fig. S3c. GA-identified motif T6 is the most representative of this group since it is connected to all other motifs in this group, and it also has the smallest peak position in the alignment score distribution (Fig. S2). Therefore, only T6 is shown with three other well separated motifs (T4, T5 and T9) in Fig. 4a of the main text as the templates to classify as-extracted clusters.

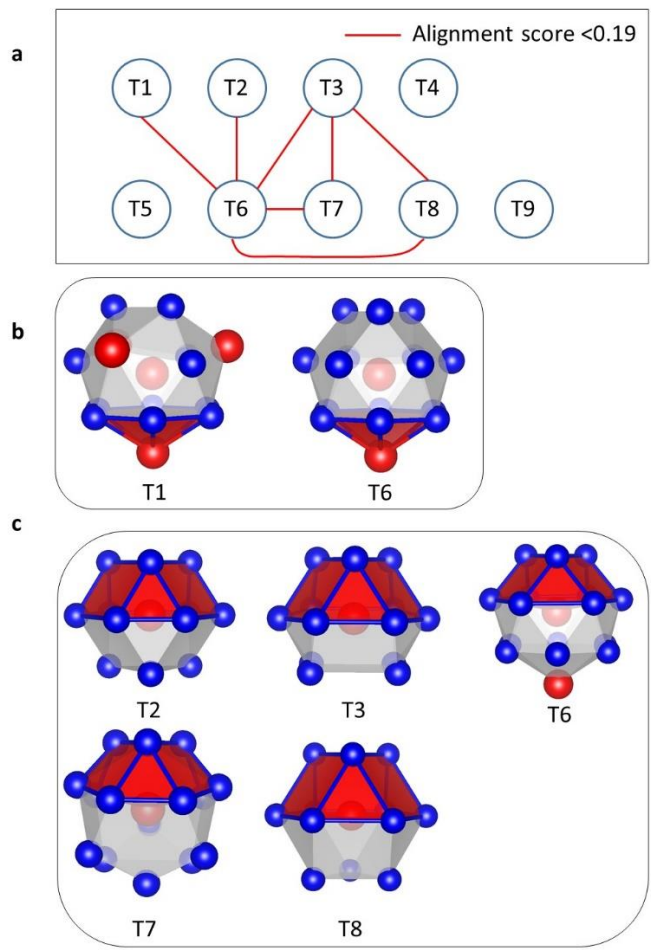

**Figure S3** Classification of the template cluster motifs by the cluster alignment algorithm. **a**, the cluster motifs are connected if their mutual alignment score is less than the cut-off value of 0.19. **b**, T1 and T6 motifs share a common Al-Sm "hexagonal pyramid". **c**, T2, T3, T6, T7 and T8 share a common "triangle and hexagon" packing of Al atoms.

**Supplementary References**